\documentclass[aps,jmp,reprint,prl]{revtex4-1}
\usepackage{graphicx}
\usepackage{amsmath}
\usepackage{amssymb}
\usepackage{dcolumn}
\usepackage{multirow}
\usepackage{hhline}
\usepackage{blindtext}
\usepackage[acronym]{glossaries}
\usepackage[utf8]{inputenc}
\newacronym{2deg}{2DEG}{two-dimensional electron gas}
\newacronym{qpc}{QPC}{quantum point contact}
\newacronym{qd}{QD}{quantum dot}
\newacronym{awg}{AWG}{arbitrary waveform generator}
\newacronym{dos}{DOS}{density of states}

\newcommand{\vpg}{\ensuremath{V_{\textnormal{PG}}}}

\newcommand{\win}{\ensuremath{W_{\textnormal{in}}}}
\newcommand{\wout}{\ensuremath{W_{\textnormal{out}}}}

\newcommand{\ef}{\ensuremath{\mu}}
\newcommand{\gout}{\ensuremath{\Gamma_\textnormal{out}}}
\newcommand{\gin}{\ensuremath{\Gamma_\textnormal{in}}}

\begin{document}
\preprint{ETH Zurich}

\title{Equilibrium free energy measurement of a confined electron driven out of equilibrium} 

\author{A. Hofmann}
\email[]{andrea.hofmann@phys.ethz.ch}
\author{V. F. Maisi}
\author{C. R\"ossler}
\author{J. Basset}
\author{T. Kr\"ahenmann}
\author{P. M\"arki}
\author{T. Ihn}
\author{K. Ensslin}
\author{C. Reichl}
\author{W. Wegscheider}
\affiliation{Laboratory for Solid State Physics, ETH Zurich}

\date{\today}

\begin{abstract}
We study out-of equilibrium properties of a quantum dot in a GaAs/AlGaAs two-dimensional 
electron gas. By means of single electron counting experiments, we measure the distribution 
of work and dissipated heat of the driven quantum dot and relate these quantities to the 
equilibrium free energy change, as it has been proposed by C.~Jarzynski 
[Phys. Rev. Lett. {\bf78}, 2690 (1997)]. We discuss the 
influence of the degeneracy of the quantized energy state on the free energy change 
as well as its relation to the tunnel rates between the dot and the reservoir.
\end{abstract}


\maketitle 

Equilibrium thermodynamics is a fundamental branch of physics providing tools to
make predictions of macroscopic many-particle systems independent of detailed 
microscopic processes governing their properties. In the recent trend towards smaller 
systems, which deviate strongly from the thermodynamic limit, fluctuations departing from 
the equilibrium state often become prominent and non-equilibrium dynamics needs to be taken 
into account. The discovery of fluctuation relations~
\cite{jarzynski_nonequilibrium_1997,crooks_entropy_1999} and their experimental tests~
\cite{collin_verification_2005,saira_test_2012,liphardt_equilibrium_2002,kung_irreversibility_2012,
carberry_fluctuations_2004,blickle_thermodynamics_2006,wang_experimental_2002,douarche_experimental_2005,
koski_distribution_2013,an_experimental_2015} 
are major steps towards understanding the evolution of small systems down 
to the atomic level. Elementary building blocks for thermodynamics on a microscopic 
level are discrete quantum states which are, for example, used for the logical 
elements of quantum bits.

Here, we study a single discrete energy level in a quantum dot coupled to a 
single thermal and electron reservoir. Driving the quantum dot out of equilibrium with
respect to the reservoir allows us 
to demonstrate experimentally the connection of an equilibrium quantity, the free energy, 
to the non-equilibrium dynamics as predicted by the Jarzynski equality~
\cite{jarzynski_nonequilibrium_1997}. The Jarzynski equality has been tested
in systems involving many energy levels and for the special case of cyclic
drive protocls~
\cite{saira_test_2012}. Other experiments exist, where the Jarzynski equality
has been used to determine the equilibrium free energy change in systems where precise
calculations thereof are difficult to obtain~
\cite{collin_verification_2005,liphardt_equilibrium_2002}.
In this experiment,
we drive a single electron occupying a fully quantized energy level and
we extract the equilibrium free energy along the full drive trajectory. 
The results are consistent with the theoretical predictions from equilibrium 
thermodynamics. We analyze the influence of a degeneracy of the discrete energy level
on the free energy.
We further discuss general limitations of fluctuation relation 
based experiments, and particularly the importance of \gls{qd} systems as testbeds
for statistical physics on the single-particle level. 

The ability to measure the free energy change in any process allows estimating the 
probability for its spontaneous occurrence. 
Even more, the free energy is a direct measure of the maximum work that may be 
extracted from a process or the minimum work which needs to be invested to activate 
it~
\cite{reichl_modern_1980}. 
A good characterization of a system in terms of the free energy is furthermore
important for experiments where drive is applied in order to cool the system, or to 
retrieve information about system properties, as well as for Szilard's engine and 
Maxwell's demon type experiments. Our results for the 
elementary single level system paves the way for understanding small systems with a 
more complex state composition, which is the case, for example, in molecular structures
coupled to a thermal bath~
\cite{gupta_experimental_2011, harris_experimental_2007, collin_verification_2005}. 

\begin{figure}[h!]
\includegraphics[width=\linewidth]{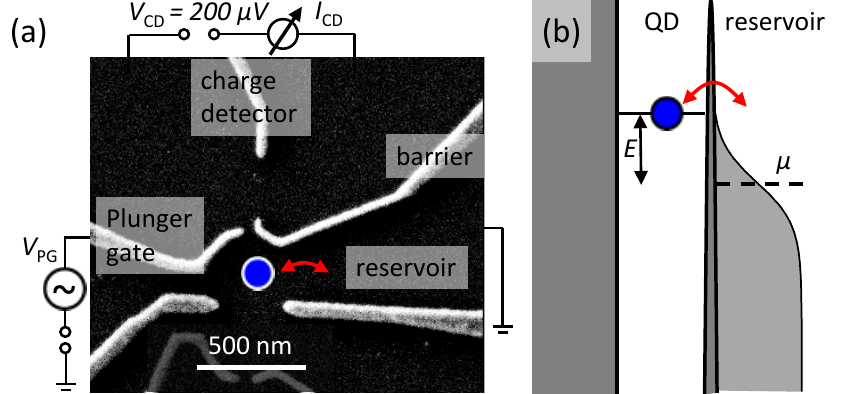}%
\caption{\label{fig:sample} (a) Scanning electron micrograph of the device
formed on a GaAs / AlGaAs heterostructure with a 2DEG 90 nm below the 
surface (dark grey area). Ti / Au gates (bright fingers) are patterned
with e-beam lithography to capacitively deplete the 2DEG. The \gls{qd} is formed in the 
place indicated by the blue circle and is tunnel 
coupled to a reservoir, as shown by the red arrow, while the other barrier is fully 
closed. An arbitrary waveform generator (AWG) is connected to the plunger 
gate to drive the \gls{qd}. (b) An energy diagram indicates a tunneling process 
between the \gls{qd} and the lead reservoir
at energy $E$ measured from the Fermi energy $\ef$.}%
\end{figure}
\begin{figure}
\includegraphics[width=\linewidth]{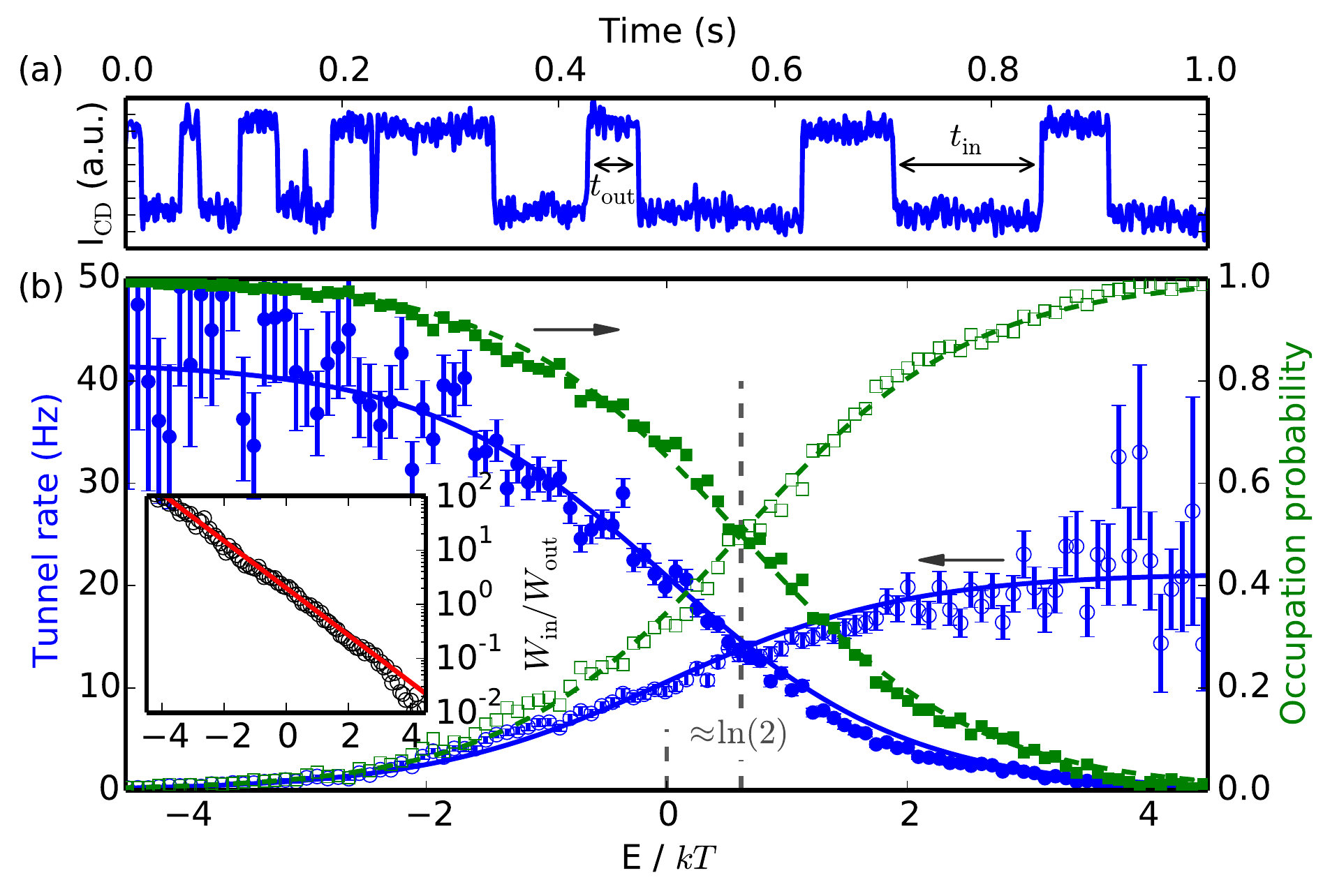}%
\caption{\label{fig:ratesoccupation} (a) A typical time trace of the charge 
detector current (I$_\textnormal{CD}$).
(b) Tunnel rates (filled blue for tunneling in and open blue dots for tunneling out) 
and relative frequency of occupation (green squares, filled for occupied and empty
for non-occupied) extracted from CD time traces of 60 s.
The abscissa is the negative plunger gate voltage measured from the Fermi energy and 
scaled by the temperature,  $T=223\ \mu$V in units of $\vpg$.
The error bars indicate statistical errors assuming Gaussian fluctiations. 
The solid blue lines are weighted least-mean-square fits 
to Eq.~(\ref{eq:rates}), from which $\gin=41.8\pm1.6$~Hz
and $\gout=21.2\pm0.6$~Hz and the temperature are determined.
The dashed green lines are fits
to Fermi functions centered around $\Delta E=k_BT\ln(2)$. Inset: The ratio of
the two tunnel rates (black circles) is plotted as a function of
energy from $\ef$, together with the exponential behaviour as expected from the detailed
balance condition (red line).}
\end{figure}

As a first experimental step we characterize the \gls{qd}-reservoir system
in thermodynamic equilibrium. A typical time trace of the charge detector current 
is shown in Fig.~\ref{fig:ratesoccupation}(a). 
In our experiment, we form a \gls{qd} in a GaAs / AlGaAs \gls{2deg}, as shown in 
Fig.~\ref{fig:sample}. The confinement can be controlled by applying negative 
voltages to top-gates. The voltage $\vpg$ applied to the plunger gate tunes
the energy level $E$ linearly. The \gls{qd} is coupled to a large contact region of the 
two-dimensional electron gas acting as the electron and heat reservoir
characterized by the temperature $T \approx 40$~mK. We measure the tunnel rates 
for the last electron~
\cite{ciorga_addition_2000, tarucha_shell_1996}
with single-electron counting techniques, utilizing a nearby quantum point contact as a 
charge detector~
\cite{schleser_time-resolved_2004, vandersypen_real-time_2004}
and find electron tunneling at frequencies of the order of 100 Hz.

The occupation probability $\tilde{f}=\sum_{i=1}^N t_\mathrm{in}^{(i)}/ \sum_{i=1}^N (t_\mathrm{in}^{(i)} + t_\mathrm{out}^{(i)})$ and 
tunneling rates $W_\mathrm{in/out} = N/\sum_{i=1}^N t_\mathrm{in/out}^{(i)}$,
with the number of tunneling events $N$, are estimated from similar time traces taken 
at the indicated plunger gate energies
\cite{schleser_time-resolved_2004,vandersypen_real-time_2004,gustavsson_counting_2006}
and presented in Fig.~\ref{fig:ratesoccupation}(b). 
The energy relaxation time for electrons injected 
from the \gls{qd} into the Fermi sea is assumed to be much shorter than 
the time scales relevant for electron tunneling between the \gls{qd} and the reservoir, and 
the chosen drive frequencies of less than $10$~Hz \cite{fujisawa_spontaneous_1998}.
Electron tunneling between the \gls{qd} and the lead is described
by Fermi's golden rule, with energy independent tunnel coupling constants $\gin$ and $\gout$
and the Fermi distribution $f(E,T)$ describing the lead occupation at an energy
$E$ measured from the Fermi energy of the lead, $\ef$. The resulting
tunneling rates are
\begin{subequations}
\label{eq:rates}
\begin{align}
 \win (E) &= \gin f(E,T) \\
 \wout (E) &= \gout \left[1-f(E,T)\right].
\end{align}
\end{subequations}
From a weighted least-mean-square fit of the measured tunnelling rates 
to Eqs.~(\ref{eq:rates}), the tunnel coupling constants, the position of the
electrochemical potential, $\ef=0$, used as the zero-energy reference, 
as well as the temperature in units of plunger gate voltage of the lead are extracted. 
The fit is shown as solid blue line in Fig.~\ref{fig:ratesoccupation}(b).
In the same figure, the solid green line is a least-mean-square fit of 
the estimated occupation probability $\tilde{f}$ to a second Fermi function, 
with the temperature and the
resonance as fit parameters. From this fit, we recover the same temperature. 
However, we find that the occupation probability 
is one half around an energy value which is offset with respect to the 
Fermi energy by $E/kT = 0.62\pm0.08$.

From the partition function of the \gls{qd}-reservoir system one can 
see that the offset of the energy where $\tilde{f}=1/2$ with respect to 
the Fermi energy is determined by 
$E = kT\ln(d)$ in the case where one electron is 
filled into an empty $d$-fold degenerate energy state. For the last electron, 
$d=2$ due to spin and hence, the occupation probability is one half at 
$E = kT\ln(2)$, as found in the experiment. 
At $E=\mu$, the probability for the energy level to be occupied is twice 
the probability for the state to be unoccupied.
As shown in the inset of Fig.~\ref{fig:ratesoccupation}(b), the system
obeys the detailed balance condition for the tunneling rates, 
$\win/\wout = d \exp(-E/kT)$, with $d=2$, as found from the measured 
$d=\gin/\gout = 1.97 \pm 0.04$.

After having characterized the \gls{qd}-lead system in equilibrium, 
our goal is to measure the equilibrium free energy change $\Delta F$
of the \gls{qd} between an initial and a final state differing by a 
value $\Delta E\propto -\vpg$ while it is driven out of equilibrium 
with respect to the reservoir by applying a time-dependent plunger
gate voltage $\vpg$. The Jarzynski equality
\cite{jarzynski_nonequilibrium_1997}
provides the necessary tools for the theoretical description of such 
an experiment. It states that for an arbitrary 
drive protocol, the performed work $\Delta W$ during the drive satisfies
\begin{equation}
\label{eq:jarzinsky}
 \langle \exp\left(-\frac{\Delta W}{kT}\right) \rangle=\exp\left(-\frac{\Delta F}{kT}\right), 
\end{equation}
where the average is taken over different repetitions of the drive
protocol and $\Delta F$ is the 
equilibrium free energy change between the initial and final state of the drive. 
As the work can be evaluated for non-equilibrium situations, 
the Jarzynski equality allows to 
determine the equilibrium quantity $\Delta F$ with a non-equilibrium measurement.

\begin{figure}[h!]
\includegraphics[width=\linewidth]{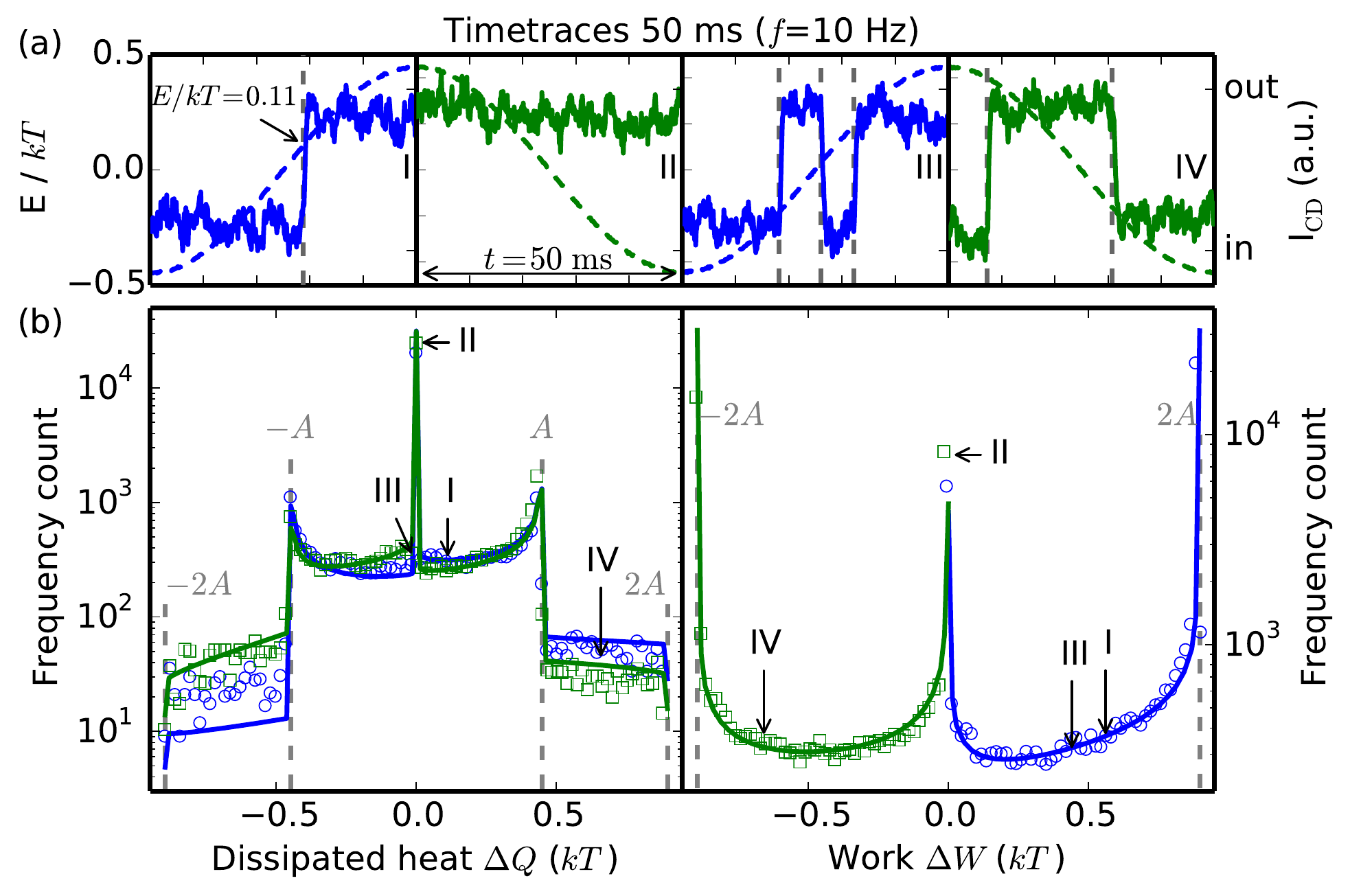}%
\caption{\label{fig:distributions} (a) Examples of drive realizations
in upward (blue) and downward (green) direction, plotted as solid lines. 
The applied drive signals are shown as dashed lines.
The driving is done by applying a voltage $\vpg \propto-E$ to the plunger gate through
an arbitrary waveform generator. From the position of the tunneling events (grey dashed lines) 
the heat and work of each 
realization are extracted. Bottom: The distribution for $\Delta Q$ (left) and $\Delta W$ 
(right) measured from approximately $20000$ realizations. Blue circles originate from 
upward and green squares from downward drive. The solid lines are 
rate equation calculations without free fitting parameters.}
\end{figure}
\begin{figure}[h!]
\includegraphics[width=\linewidth]{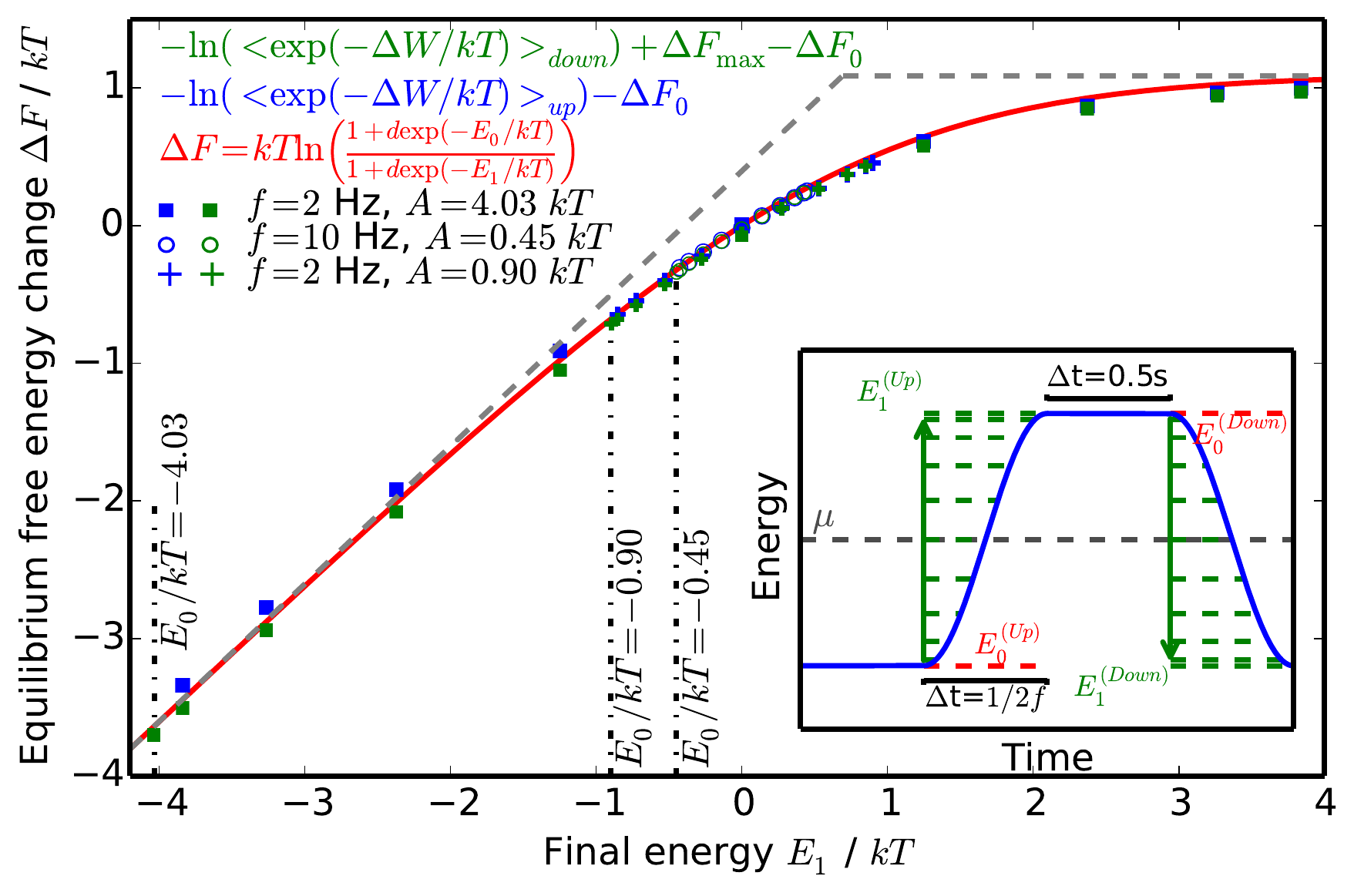}%
\caption{\label{fig:FreeEnergy} The free energy of the \gls{qd} during the drive,
obtained with non-equilibrium measurement, measured with respect
to the Fermi energy. Experimental results 
are plotted in green for the downward and in blue for 
the upward drive direction. Different markers denote independent experiments
with different drive parameters.
The theory curve, Eq.~\ref{eq:freenergy}, is plotted as solid red line,
for $d=2$.
Dashed grey lines indicate the expected linear increase at low $E_1$ 
as well as the saturation value for $E_1>E_F$.
Inset: One full period of the drive is plotted in blue. A waiting 
time of 0.5~s ensures thermal equilibration of the initial state. Green dashed
lines indicate the values of $E_1$ for which we extract $\Delta F$ 
from the experiment.}
\end{figure}

In our experiment, we apply a sinusoidal voltage ramp $\vpg(t)$ to 
the plunger gate, which changes the energy of the electronic state in the 
\gls{qd} from $E_0$ to $E_1$ and performs work on it as long as it is
occupied [see Fig.~\ref{fig:sample}(b)].
The drive protocol, as shown in the inset of Fig.~\ref{fig:FreeEnergy},
is characterized by the frequency $f$ and amplitude $A$ of the sinusoidal
part of the protocol.
Together, $A$ and $f$ determine the maximum steepness of the ramp and the initial 
($E_0=-A$) and final ($E_1$) values of the \gls{qd} energy level.
We distinguish two drive directions, upward ($E_1 > E_0$) and downward ($E_0 > E_1$),
and employ a waiting time of $\Delta t =0.5$~s for equlibration in-between. 
The respective initial and final states,
$E_0^{(Up,Down)}$ and $E_1^{(Up,Down)}$, are indicated in the figure.
Typical time-traces showing the charge-detector signal for four
individual voltage ramps are shown in Fig.~\ref{fig:distributions}(a).
In most cases, as shown on panel~I, when the energy level of the QD
starts well below the Fermi level and is raised above the Fermi level, 
the electron leaves the \gls{qd}. If the drive frequency is large compared
to the tunneling rates, realizations without tunneling events become probable, 
as shown in panel~II.
For each realization, we determine the work $\Delta W$ perfomed on 
the \gls{qd}. When driving an occupied energy 
level of the \gls{qd}, work is performed on the electron by the voltage 
source amounting to
\begin{align}
 \label{eq:work}
 \Delta W = \int_{E_0}^{E_1} n(E) dE,
\end{align}
where $n(E)\in \{0,1\}$ denotes the occupancy of the \gls{qd}, which
is directly measured by the charge detector signal. For example, in 
realization~I, with the drive applied symmetrically around zero 
and $A = 0.45\ kT$, the work performed on the electron in 
the \gls{qd} is $\Delta W = (0.45 + 0.11)\ kT = 0.56\ kT$.
Note that the maximum amount of work that can be performed on or by the QD system 
is given by $\pm 2A$.

In addition, for each realization we determine the heat $\Delta Q$ dissipated in the 
contact from the energies $E_i$ of the \gls{qd} state at which tunneling events
occur. Each tunneling process contributes $\Delta Q_i = E_i$ for tunneling out
and $\Delta Q_i = -E_i$ for tunneling into the \gls{qd}. For example, in 
realization~I, the electron tunnels at $0.11\ kT$ above $\ef$, hence 
$Q=0.11\ kT$ due to the relaxation of the electron in the lead.
If more than one tunneling event occurs during a single realization,
as shown in panels~III and IV, the individual processes are added up,
\begin{align}
 \Delta Q=\sum_i \Delta Q_i = \sum_i E_i s_i,
\end{align}
where $s_i$ distinguishes between tunneling-out processes, $s_i=1$, and 
tunneling-in processes, $s_i=-1$. The maximum amount of heat which can be
dissipated in the lead during a single drive realization amounts 
to $\pm 2A$.

With the analysis described above, we now determine the relative frequency distributions 
of $\Delta W$ and $\Delta Q$ as shown in Fig.~\ref{fig:distributions}(b).
The distributions for driving upwards
are plotted as blue circles, while green squares are used for the distributions resulting 
from driving downwards.
With the tunnel rates given in Fig.~\ref{fig:ratesoccupation}(b),
$\win = 42$~Hz and $\wout=21$~Hz, and a drive frequency of $f=10$~Hz, realizations as shown 
in panel~II of Fig.~\ref{fig:distributions}(b), i.e. without tunnel events
and hence zero dissipated heat, are very probable, leading to a prominent delta-peak at 
$\Delta Q=0$ in Fig.~\ref{fig:distributions}(b). 
In these realizations, no work is done on the \gls{qd} if the level stays empty during 
driving, leading to a peak at $\Delta W=0$ in Fig.~\ref{fig:distributions}(b). 
If it is occupied, the work 
equals plus or minus twice the drive amplitude for upward or downward drive 
direction, respectively. Hence, we observe two additional peaks, at $\Delta W = \pm2A$,
also corresponding to the single delta-peak at zero dissipation. From 
Fig.~\ref{fig:distributions}(b), it is apparent that the relative frequency for 
$|Q|\leq A$  is much higher than for $|Q|>A$. 
This is due to the fact that the maximum dissipation in a single 
realization depends on the occupancy of the initial and final state. 
For example, when driving an energy level downwards, the dissipated heat 
can only exceed $A$ if the state has been occupied in the beginning, 
as in realization~IV. Contrary, $\Delta Q < A$ is only possible for downdrives 
if the final state is the less probable unoccupied state. 
Since the \gls{qd} is thermally equilibrated by the waiting time before each 
drive, the occupancy of the initial state is given 
by the Fermi distribution and realizations with large $|Q|$ are suppressed. 
This effect, together with a comparably fast sinusoidal
drive, also leads to the peaks found at $Q=\pm A$: they are mostly due to 
realizations starting in the more probable initial state and having a single
tunnel event at the very end of the drive. Generally, the sharp features found
in the probability distributions shown
here are the signatures of non-equilibrium dynamics. For slower drives, the probability
for intermediate dissipation $|Q|\sim0$ increases and we find more Gaussian
shaped distributions centered around zero dissipated heat (data not shown). 
This corresponds to the adiabatic limit, where a Gaussian shape with a width 
given by the fluctuation-dissipation theorem is expected~
\cite{nyquist_thermal_1928}, as discussed in more detail in Ref.~
\onlinecite{saira_test_2012, douarche_experimental_2005}
The detailed shape of the distributions depends on the relative amplitude $A/kT$ and 
frequency $f/\Gamma_{in,out}$ of the sinusoidal drive. For example, the distributions 
for work and dissipated heat are asymmetric with respect to the two drive directions 
because of the different rates for tunneling in and out of the \gls{qd}, as discussed
above. Using the standard rate equation approach
~\cite{saira_test_2012} with parameter values extracted from the measurements of
Fig.~\ref{fig:ratesoccupation}(b), we calculate the probability distributions for the 
work and the dissipated heat in each drive direction and find very good agreement 
with the experimental data.

Utilizing Eq.~(\ref{eq:jarzinsky}), we now determine the equilibrium free energy 
change of the \gls{qd} from the statistics of performed work. The prerequisite 
that the initial state must be thermally equilibrated \cite{jarzynski_nonequilibrium_1997} 
is fulfilled by the waiting time. We determine the free energy change between the fixed initial
and any freely chosen final state within the drive by analyzing $\Delta W$ as described 
above. As a result we get the free energy difference of the initial state and the chosen 
state within the drive. By varying the final state in the analysis, the change in 
free energy is determined along the full drive trajectory, as shown in
Fig.~\ref{fig:FreeEnergy}. Different markers are used for the three measurements
performed with the same gate configuration and \gls{qd} resonance but with 
different values for $A$ and $f$, as indicated in the figure.
Again, blue color denotes upward and green denotes downward drive direction. 
Each of the six drive protocols gives an independent estimate for $\Delta F$.
From the extracted $\Delta W$ we also find that the average work $\langle W \rangle$
always exceeds the free energy change, as is expected for a system which is
driven out of equilibrium.

Theoretically, the free energy is calculated 
from the difference of the partition sums of the initial ($i=0$) and final ($i=1$) states
\cite{reichl_modern_1980},
\begin{align}
\label{eq:freenergy}
 \Delta F = kT \ln\left(\frac{Z_0}{Z_1}\right),\ 
 Z_i = 1 + d \exp\left(\frac{-E_i}{kT}\right)
\end{align}
The theoretical curve is plotted as a solid red line in Fig.~\ref{fig:FreeEnergy}
for $d=2$, which is the value obtained from the equlibrium
characterization above. Compatible with theory, we observe a linear 
increase of the free energy in the first part of the drive, where the electron is 
lifted to an energy level still well
below the Fermi energy. The increase in $\Delta F$ slows down around the Fermi 
energy, as the probability of the \gls{qd} to be occupied decreases. At energies
well above the Fermi level, the occupation probability approaches zero and
the free energy saturates. The six independent measurements of $\Delta F$ agree
well with the theoretical value along the whole drive trajectory.

Small systematic deviations from the theoretical value of $\Delta F$ are found only for the data
resulting from updrives at $f=2$~Hz and $A=4.03\ kT$. Experimental 
limitations are given by the stability of the sample as well as
the bandwidth of the set up. The bandwidth limits the accuracy in the determination
of the exact tunnel time in drive schemes where large amplitude and frequency 
are combined. These errors lead to uncertainties in $\Delta W$ and thereby also
in $\Delta F$. The sample stability, on the other hand, sets a limitation to the
number of repetitions which can be performed in a given configuration of the \gls{qd}.
Most importantly, the drive must be applied around the chosen operation point,
equally for every repetition. Energy drifts in time therefore lead to systematic errors.
Although we observe stable sample configurations over many hours, we implemented a 
feedback mechanism in order to align $E=0$ to the Fermi level $\mu$ every quarter 
hour. Offsets are taken into account in the analysis.

As demonstrated in Fig.~\ref{fig:FreeEnergy}, we have found good agreement 
between the experimental 
determination and theoretical expectation of the free energy
change in our well-characterized system.  This
supports the idea that the free energy can be measured at
any point of the drive in a system far out of equilibrium, independent
of the drive parameters, by utilizing the Jarzynski
equality. In the original work by C. Jarzinsky, an experimental test of equation
\ref{eq:jarzinsky} is suggested in a nanoscale system weakly coupled to a
reservoir, where the fluctuation of the work is less then $kT$ and which
evolves deterministically under its Hamiltonian~
\cite{jarzynski_nonequilibrium_1997}. Even though the phase space evolution of 
the \gls{qd} in our case is intrinsically non-deterministic due to 
tunneling events and the fluctuations are of the order of $kT$
drive amplitudes, we find Eq.~\ref{eq:jarzinsky} to be valid and
the \gls{qd} system to be a very suited testbed for the Jarzinsky equality.

As known from equilibrium thermodynamics, the free
energy is a state variable and therefore independent of
the trajectory. When the system is driven out of equilibrium, such trajectory-independent 
quantities are not generally available and the situation becomes more complex. In this 
work we have shown that an equilibrium quantity, the free energy of a two-fold 
degenerate single electron state, can be measured with an experiment driving the 
system far from equilibrium. Our results show that thermodynamic quantities such as 
work and dissipation can be understood on the level of individual quantum states and 
in nonequilibrium. The good agreement with theoretical calculations proves that the 
simple system we have used suits well for studying more complex phenomena out of 
equilibrium. As a direct consequence of the presented work, we expect the relation 
between thermodynamics and information theory as well as influences of dissipation 
on future fast drives of qubits to be studied in quantum dot systems.

\begin{acknowledgments}
We want to thank the SNF and QSIT for providing the funding which enabled this work.
\end{acknowledgments}
\bibliography{mybib}
\bibliographystyle{apsrev4-1}
\end{document}